\documentclass[twocolumn,footinbib,prb,showkeys,amsmath,amssymb]{revtex4}

\usepackage{graphicx}
\usepackage{dcolumn}
\usepackage{bm}

\begin{document}

\preprint{}

\title{Low Dimensional Lattice Diffusion in Solids Investigated by Nuclear Spin Echo Measurements}

\author{Naoki Asakawa}
\thanks{Corresponding Author}
\email{nasakawa@bio.titech.ac.jp}
\affiliation{Department of Biomolecular Engineering, Tokyo Institute of Technology.  4259 Nagatsuta-cho, Midori-ku, Yokohama, Kanagawa 226-8501, Japan}%

\author{Kiyohiko Matsubara}
\affiliation{Department of Biomolecular Engineering, Tokyo Institute of Technology.  4259 Nagatsuta-cho, Midori-ku, Yokohama, Kanagawa 226-8501, Japan}

\author{Yoshio Inoue}
\affiliation{Department of Biomolecular Engineering, Tokyo Institute of Technology. 4259 Nagatsuta-cho, Midori-ku, Yokohama, Kanagawa 226-8501, Japan}

\date{\today}

\begin{abstract}
Lattice diffusion and internal local magnetic field gradients in solids
are investigated by numerical simulation of nuclear mangetic resonance(NMR) spin echo experiments.
The Fourier-spectrum method is employed
in order to solve the Bloch-Torrey equation
with arbitrary magnetic field gradients
in one- and two-dimensional lattice restrictions.
\end{abstract}

\maketitle

\section{Introduction}
Transport phenomena are often encountered 
in solutions, fluids, gels, and
quasi particles associated with elementary excitations in solids,
and so on.
In particular,
low dimensional transport phenomena are received much attraction in various research fields 
of electronic devices such as polymeric conducting materials\cite{PA}, 
quasi one-dimensional organic conductors~\cite{q1Doc}
and rechargeable batteries~\cite{Conradi1,Conradi2}.
In these matarials or devices,
low dimensional diffusive or hopping motion 
such as charged ions, defects, quasi-particles(e.g., exciton, polaron, conformon)
in their crystalline lattices 
plays an important role in origins of their physical properties.
Some of such molecular motions in crystalline solids are called
"lattice diffusion," 
which is defined as transport phenomena of (quasi-)particles                
of which positions are restricted in lattice points in crystals. 
The motion often leads to motional averaging of local magnetic fields  
 in solids.
Nuclear magnetic resonance(NMR) spectroscopy is 
one of promising nondestructive methods
for investigating these phenomena.
Several attempts have recently been reported to make NMR measurements
of ultra slow diffusion in solids\cite{Kimmich,Chang,Ailion}.

Among these, the pulse field gradient(PFG) technique with strong magnetic field gradient(MFG) has been given much attention for detection of such diffusions.
However, the PFG technique involves some problems in cases, which are
(i) attenuation of sensitivity in association of application of PFG,
(ii) a lack of magnitude of PFG for ultra slow diffusion measurements,
and (iii) the case that the system has large internal MFG($G_{\rm i}$) compared to external PFG and/or large heterogeneous MFG.

The problem (i) and (ii) are often pronounced in molecular diffusion measurements.
Since the problem (iii) is not problematic in cases of systems 
with large diffusion coefficient, it has often been ignored.
In such cases, the internal MGF is thought to be averaged to zero, 
and then effects of the internal MFG will be small or negligible.
However, in solids, the diffusion coefficient can be quite small and 
a local internal MFG, which in some cases heterogeneous and/or anisotropic, 
can be remained\cite{Ailion}.

In such situations, an effective MFG is no longer the same as the external MFG
($G_{\rm e}$),
which is likely to mask the accurate diffusion coefficient.
Even in the case that the system has large local MFG,
it will not be problematic 
if the root mean squared displacement by diffusion($\sqrt{Dt}$) is much larger than 
the periodicity of the local MFG,
because the local MFG is expected to be averaged out during diffusion.
On the contrary,
in the case of systems with small diffusion coefficient,
the local MFG can be remained and spin echo trains of magnetic resonance 
can be further attenuated by the local MFG as well as by the external MFG.

So far,
there are several kinds of possible approaches to cancel the effect of local MFG,
that is to set up the experimental condition of 
$G_{\rm e} \gg G_{\rm i}$ 
by production of large external MFG by
an anti-Helmholtz superconducting magnet
or a fringe field of a superconducting magnet.
By using anti-Helmholtz superconducting magnet,
one can obtain a MFG up to 200 T$\cdot$m$^{-1}$\cite{Chang}.
On the other hands, 
by using superconducting fringe field, one can obtain up to 60 T$\cdot$m$^{-1}$\cite{Kimmich}.
There are the other approaches to cancel local MFG 
by using bipolar gradients\cite{bipolar1,bipolar2,bipolar3,bipolar4} or rotary echos\cite{Kimmich-review}.
These methods are based on the idea of cancelation of unwanted local MFG. 
On the contrary, 
we are interested in local MFG, which is thought to be 
characteristic to each systems particularly to solids, and 
it should be closely related to the molecular and electronic structure. 

There are a number of literatures available to calculate spin echos 
in restricted geometries~\cite{Fukushima,Kenkre,Callaghan,Barzykin,Sen}.
Most of these researches are devoted to describe pulse gradient spin echo(PGSE) experiments.
The spin echo experiment under a static field gradient is a kind of niche applications, 
and its theoretical treatment, namely, solvation  of Bloch-Torrey equation~\cite{BTE}, is still lacking.
Up to now, Axelrod and Sen have shown a solution of one dimensional Bloch-Torrey equation under the linear gradient condition
by using an eigen mode technique\cite{Sen}.  
In this Letter, 
we shall extend their treatment 
to spin echo experiments with two-dimensional lattice diffusion under arbitrary magnetic field gradient shape.  
Although the general theory of spectral density function for two-dimensional lattice diffusion is developed~\cite{Sholl},
effects of 2D diffusion on nuclear spin echo experiments have not been investigated so far.

The method developed here will be useful to investigate transport phenomena in low dimensional solids,
such as 
movement of lattice defects(including soliton in disordered solids), 
dynamics of elementary excitation(including exciton and polaron),
transport of charge carriers, 
and so on.
Further, 
recent developments on magnetic resonance force microscopy(MRFM) show
that one can perform {\it in situ} imaging with a scale of several tens of nanometers,
where ultra high magnetic field gradients of over 10$^3 \sim 10^4$ T/m are generally used~\cite{Garbini1,Garbini2}.
Up to now, 
there are no publications available to obtain spin-echo measurements 
under a condition of such a huge MFG, which can be comparable to local magnetic field gradient with atomic scale.
The theory presented in this Letter will be useful for analyses of such experiments as well.

\section{Theoretical}
The method presented here is based on the numerical solution of 
the Bloch-Torrey equation by the Fourier-spectrum method\cite{FS}.
Axelrod and Sen have proposed the methodology for calculating 
amplitudes of spin echos for a nucleus diffusing under a linearly 
inhomogeneous magnetic field condition\cite{Sen}.
Although they have shown the solution with a linear magnetic field gradient 
by an eigen mode expansion method, 
the details of the numerical procedure was not shown.
In this article,
we shall demonstrate a mathematical formulation of solvation of the Bloch-Torrey equation by
the Fourier-spectrum method and
explore spin echo amplitudes under various magnetic field gradients.

So far,
we have attempted to solve the differential equation of diffusion equation 
in the real space by Crank-Nicolson method\cite{CN}.
However, 
we have suffered from the underflow of variables due to iterative  
multiplication in many times, of the time evolution operator 
for the spin echo experient. 
Furthermore, while the Crank-Nicolson method preserves the absolute stability against the size of each step in the real space,
we observed the strong dependence of spin echo amplitudes 
on magnitude of interval of time step.

Thus, these problems lead us to motivation for numerical simulation 
in the Fourier space concerning coordinates in real space.

Below we shall describe the mathematical formulation for Bloch-Torrey equation
with magnetic field gradient of an arbitrary wave form. 

\subsection{Solution for 1D Bloch-Torrey Equation with arbitrary magnetic field gradient}
The Bloch-Torrey equation is shown in the following:
\begin{eqnarray}
\frac{\partial M}{\partial t} &=& \tilde{D_0} \frac{\partial^2 M}{\partial \tilde{x}^2} -i \tilde{\gamma} f(\tilde{x}) M, \label{Bloch-Torrey}
\end{eqnarray}
where the tilde denotes dimensionless parameter in order the problem to be 
treated as general. 
The effects of longitudinal relaxation($T_1$) and intrinsic transverse
relaxation($T_2$) in original Bloch-Torrey equation was neglected 
in the calculation.
But the effect must of course be taken into account in practical applications.
Dimensionless diffusion coefficient($\tilde{D}_0$), 
dimensionless gyromagnetic ratio($\tilde{\gamma}$), 
and one-dimensional coordinate($\tilde{x}$) are defined as 
\begin{eqnarray}
\tilde{D}_0 &=& \frac{D_0 \tau}{L_s}\\
\tilde{\gamma} &=& \tau g \gamma L_s\\
\tilde{x} &=& \frac{x}{L_s}. 
\end{eqnarray}
$f(\tilde{x})$ is a functional form of 1D MFG.

Let us think the expansion of the Eq.(\ref{Bloch-Torrey}) by a series of orthogonal functions, $\Psi_k(\tilde{x})$.
\begin{eqnarray}
{\mathcal F}[\frac{\partial M}{\partial t}] &=& \tilde{D}_0 {\mathcal F}[\frac{\partial^2 M}{\partial \tilde{x}^2}] -i \tilde{\gamma} {\mathcal F}[f(\tilde{x})M].\label{BTk}
\end{eqnarray}
Here, we used linearity of orthogonal function transformation.
The second term of the right hand side of Eq.(\ref{BTk}) is obtained by 
convolution: 
\begin{eqnarray}
{\mathcal F}[fg] &=& {\mathcal F}[f]{\bf *}{\mathcal F}[g].
\end{eqnarray}
Thus, one can write down the practical form of Eq.(\ref{Bloch-Torrey}) like the following:
\begin{eqnarray}
\frac{dM_k(t)}{dt} &=& -(\pi k)^2 \tilde{D}_0 M_k(t) 
  - i \tilde{\gamma} \sum^{N-1}_{p=0}V_{k-p} M_p(t), \label{conv}
\end{eqnarray}
Here,
\begin{eqnarray}
M_k(t) &=& {\mathcal F}[M] = \int^{1/2}_{-1/2} M(\tilde{x},t) \Psi_k(\tilde{x}) d\tilde{x}.
\end{eqnarray}
$V_k$ is a k-th coefficient for orthogonal function transform(for example,
Fourier transform) of $\tilde{x}$:
\begin{eqnarray}
V_k &=& \int^{1/2}_{-1/2} f(\tilde{x}) \Psi_k(\tilde{x}) d\tilde{x}.
\end{eqnarray}
Further, the first term of the right hand side of Eq.(\ref{conv})
can be obtained by using the fact that second derivative in real space corresponds to
the multiplication with $k^2$.

\begin{eqnarray}
\frac{d M_k(t)}{dt} &=& \left( {\Hat W} - i \tilde{\gamma} {\Hat B} \right) M_k(t) \label{Master}
\end{eqnarray}
Here, 
\begin{eqnarray}
{\Hat W}_{kl} &=& \left\{
\begin{array}{ll}
-(\pi \cdot k)^{2} \tilde{D}_0 \mbox{\ \ if $k = l$}\\
0 \mbox{\ \ if $k\neq l$}
\end{array}
\right.
\end{eqnarray}
\begin{eqnarray}
{\Hat B}_{kl} = V_{k-l}
\end{eqnarray}

Let us think a basis set with taking account of the boundary condition 
of $\tilde{x}$:[-1/2:1/2].
Shown below is 
the eigen modes using sinusoidal(only cosine components) orthogonal functions
for the Laplacian, $\tilde{\Delta_0} = - \tilde{\nabla}^2$,
of the one dimensional diffusion equation,
\begin{eqnarray}
\Psi_k(\tilde{x}) &=& C_k \cos ( \pi k (\tilde{x}+\frac{1}{2}) ),
\end{eqnarray}
where 
\begin{eqnarray}
C_k &=& \left\{
\begin{array}{ll}
1 & \mbox{if $ k = 0 $}\\
\sqrt{2} & \mbox{if $ k \neq 0 $}
\end{array}
\right.
\end{eqnarray}
and corresponding eigen values are $\tilde{\lambda}_k = (\pi k)^2$. 

Since it is unnecessary to think the time ordering 
during the echo time, $\tau$, 
in transverse relaxation time measurements,
one can take $\tau$ as one step.

The time evolution operator for the echo time, $\tau$, can be defined as
\begin{eqnarray}
U_+ = e^{({\Hat W} - i \tilde{\gamma} {\Hat B})\tau }.
\end{eqnarray}
Thus,
the propagator for the two pulse Hahn echo~\cite{Hahn} can be express as 
\begin{eqnarray}
U_{2\tau} = U_- U_+ = (U_+)^* U_+.
\end{eqnarray}
Similarly, that for the second echo of Carr-Purcell Meiboom-Gill(CPMG)~\cite{CP,CPMG} can be expressed as 
\begin{eqnarray}
U_{4\tau} = U_+ U_- U_- U_+,
\end{eqnarray}
and that for n-th echo of CPMG, 
\begin{eqnarray}
U_{2n\tau} = U_{2\tau}^{{\rm mod}(n,2)} U_{4\tau}^{{\rm floor}(n,2)}.
\end{eqnarray}
Therefore,
The k-th Fourier component of the magnetization at a time, $t=2 n \tau$, can be calculated by 
\begin{eqnarray}
M_k(t) &=& U_{2n\tau} M_k(0) \label{Mkt}.
\end{eqnarray}
Here, the initial magnetization,  $M_k(0)$, can be obtained 
by using the initial condition of $\phi(\tilde{x})$:
\begin{eqnarray}
\sum^{\infty}_{k=0} M_k(0) \Psi_k(\tilde{x}) 
&=& \sum^{\infty}_{k=0} M_k(0) \cos(\pi k (\tilde{x}+\frac{1}{2})) \nonumber \\
&\equiv& \phi(\tilde{x}).
\end{eqnarray}
The Fourier component, $M_k(0)$, of the initial magnetization, can be
calculated by inverse Fourier transform of $\phi(\tilde{x})$.
For example, 
in the case of homogeneous spin density, namely, 
in the case of the initial condition of $\phi(\tilde{x}) = 1$(constant),
all the components are zero except k=0 
since the inverse Fourier transform is a $\delta$-function.
Substituting the obtained $M_k(0)$ into Eq.(\ref{Mkt}),
the k-th magnetization, $M_k(t)$, at a time, $t(=2n\tau)$,
can be computed.
Finally, 
one can calculate the magnetization, $M(t)$ by inverse Fourier transform of 
$M_k(t)$ and the summation of each components which correspond to
each components in real space;

\begin{eqnarray}
M(t) &=& \sum_{\tilde{x}=-1/2}^{1/2} {\mathcal F}^{-1}[M_k(t)].
\end{eqnarray} 
All the figures in this article are plots of the relaxation exponent, $\log \mid \frac{M(0)}{M(t= 2 n \tau)} \mid /n$, as a function of $\tilde{D}_0$. 


\subsection{Two-dimensional Diffusion}
We extended the 1D Bloch-Torrey equation to the case of anisotropic 2D lattice diffusion. 
For anisotropic 2D diffusion,
the Bloch-Torrey equation is expressed as follows.
\begin{eqnarray}
\frac{\partial M}{\partial t} &=& \tilde{D}_{\tilde{x}0}\frac{\partial^2 M}{\partial \tilde{x}^2
} + \tilde{D}_{\tilde{y}0}\frac{\partial^2 M}{\partial \tilde{y}^2} -i \tilde{\gamma} f(\tilde{x
},\tilde{y}) M.\label{2D_BT}
\end{eqnarray}
The solution can be written as
\begin{eqnarray} 
M(t) &=& U_{2n\tau} M(0),
\end{eqnarray} 
where
\begin{eqnarray} 
U_{2n\tau} &=& \exp\Bigl(-\int ^{2n\tau}_0 (-D_0\nabla ^2 -i\gamma \Hat{f}(\tilde{x}, \tilde{y})) dt \Bigr) 
\end{eqnarray} 

Now we perform Fourier transform using the series of 
two dimensional orthogonal functions, $\Psi _{mn}(\tilde{x}, \tilde{y})$,  
\begin{eqnarray}
\Psi _{mn}(\tilde{x}, \tilde{y}) &=& C_{mn} \Psi _{m}(\tilde{x}) \Psi _{n}(\tilde{y}) \nonumber \\ 
&=& C_{mn} \cos \bigl(\pi m (\tilde{x} + \frac{1}{2})\bigr)\cos\bigl(\pi n(\tilde{y} + \frac{1}{2})\bigr) \nonumber\\
\end{eqnarray}

$C_{mn}$ is a normalization constant and $m$, $n$ are eigen mode values.
\begin{equation}
C_{mn} = \begin{cases}
	1 &        (m = n = 0) \\
	\sqrt{2} & (m = 0  \hspace{0.25cm} {\rm or}  \hspace{0.25cm} n = 0) \\
	2 &        (m \neq 0 \hspace{0.25cm} {\rm and}  \hspace{0.25cm} n \neq 0) \\
      \end{cases} 
\end{equation}

Let us consider the space Fourier transform of Eq.(\ref{2D_BT}):
\begin{eqnarray}
\mathcal{F}[M_t] &=& D_0\mathcal{F}[\nabla ^2 M] - i\gamma\mathcal{F}[f(\tilde{x}, \tilde{y}) M]. \label{2D_BT_Fourier}
\end{eqnarray}

Again, 
$\mathcal{F}[M]$ stands for the 2D Fourier transform of magnetization $M$,
\begin{eqnarray}
\Hat{M}_{mn}(t) &=& \mathcal{F}[M] \nonumber\\
&=& \iint ^{\frac{1}{2}}_{-\frac{1}{2}} M(\tilde{x}, \tilde{y}, t)\Psi _{mn}(\tilde{x},\tilde{y}) d\tilde{x}d\tilde{y}
\end{eqnarray}

Here, $\Hat{M}$ represents the magnetization in the Fourier space.
We begin by the procedure for calculating $\mathcal{F}[\nabla ^2 M]$ of Eq.(\ref{2D_BT_Fourier}). 
\begin{eqnarray}
\mathcal{F}[\nabla ^2 M] = \mathcal{F}[M_{\tilde{x}\tilde{x}}] + \mathcal{F}[M_{\tilde{y}\tilde{y}}] \hspace{3.25cm} 
\end{eqnarray}

For the second-order derivative of the $\tilde{y}$ component,

\begin{eqnarray}
\mathcal{F}[M_{\tilde{y}\tilde{y}}] \hspace{7.5cm}\nonumber \\
=
\begin{pmatrix}
-(\pi\cdot 0)^2 M_{00} & -(\pi\cdot 1)^2 M_{01} & \ldots & -(\pi\cdot n)^2 M_{0n}\\ 
-(\pi\cdot 0)^2 M_{10} & -(\pi\cdot 1)^2 M_{11} & \ldots & -(\pi\cdot n)^2 M_{1n}\\ 
\vdots & \vdots &\ddots & \vdots \\
-(\pi\cdot 0)^2 M_{m0} & -(\pi\cdot 1)^2 M_{m1} & \ldots & -(\pi\cdot n)^2 M_{mn}\\
\end{pmatrix}
\nonumber
\end{eqnarray}
\begin{eqnarray}
&=&
\begin{pmatrix}
M_{00} & M_{01}  & \ldots & M_{0n}\\ 
M_{10} & M_{11}  & \ldots & M_{1n}\\ 
\vdots & \vdots & \ddots & \vdots \\
M_{m0} & M_{m1} & \ldots & M_{mn} \\
\end{pmatrix}
\nonumber \\
&\times&
\begin{pmatrix}
-(\pi\cdot 0)^2 &       0          & \ldots &    0     \\ 
       0        & -(\pi\cdot 1)^2  & \ldots &    0      \\ 
    \vdots      &      \vdots      & \ddots & \vdots \\
       0        &       0          & \ldots & -(\pi\cdot n)^2 \\
\end{pmatrix}
\nonumber \\
&=& \Hat{M} \Hat{W}_y
\end{eqnarray}

Similarly, for the $\tilde{x}$ component, $\mathcal{F}[M_{\tilde{x}\tilde{x}}] = \Hat{W}_x \Hat{M}$,

where,
\begin{equation}
\Hat{W}_x = 
\begin{pmatrix}
-(\pi\cdot 0)^2 &       0          & \ldots &    0     \\ 
       0        & -(\pi\cdot 1)^2  & \ldots &    0      \\ 
    \vdots      &      \vdots      & \ddots & \vdots \\
       0        &       0          & \ldots & -(\pi\cdot m)^2 \\
\end{pmatrix}
\end{equation}

Thus,
\begin{eqnarray}
\mathcal{F}[M_{\tilde{x}\tilde{x}}] + \mathcal{F}[M_{\tilde{y}\tilde{y}}] = \Hat{W}_x \Hat{M} + \Hat{M} \Hat{W}_y
\end{eqnarray}

When $m = n$, both $\Hat{W}_x$ and $\Hat{W}_y$ are same diagonal matrices each other, 
so we can rewrite,  
\begin{equation}
\mathcal{F}[M_{\tilde{x}\tilde{x}}] + \mathcal{F}[M_{\tilde{y}\tilde{y}}] = \Hat{W}_x \Hat{M} + \Hat{M} \Hat{W}_x.
\end{equation}

Now we introduce a Laplacian superoperator $\Hat{\Hat{W}}$ 
based on commutator superoperators~\cite{Ernst}.
\begin{eqnarray}
\Hat{\Hat{W}} \Hat{M} &=& \Hat{W}_x \Hat{M} + \Hat{M} \Hat{W}_x \nonumber \\
&=& \Hat{W}_x \Hat{M} \times E + E \times \Hat{M} \Hat{W}_x,
\end{eqnarray}
where $E$ represents the unity operator. This leads to the matrix representation,
\begin{eqnarray}
\Hat{\Hat{W}} = \Hat{W}_x \otimes \Hat{E} + E \otimes \tilde{W}_x
\end{eqnarray}
where $\tilde{W}_x$ represents the transposed matrix of ${\Hat W}_x$. 
$\otimes$ denotes direct product of matrices.

In the practical calculation, we assume the square lattice and adopt the identical number of eigen modes for $\tilde{x}$ and $\tilde{y}$ dimensions($m = n$), 
and $\tilde{E} = E$.    
\begin{eqnarray}
\mathcal{F}[\nabla ^2 M] = \Hat{\Hat{W}} \Hat{M}
\end{eqnarray}

Now we go on to show the calculation of the second term, ${\cal F}[f(\tilde{x}, \tilde{y}){\mathbf M}]$, in Eq.(\ref{2D_BT_Fourier}).
The Fourier transformation of this term can be rewritten using a convolution, 
\begin{eqnarray}
\mathcal{F}[f(\tilde{x}, \tilde{y}) M] &=& \mathcal{F}[f(\tilde{x}, \tilde{y})] \ast \mathcal{F}[M]  \nonumber \\
&=& \sum _{p} \sum _{q} V_{m-p, n-q}M_{p, q}, \label{2Dconv}
\end{eqnarray}
where $V_{m, n}$ represents $(m, n)$-$th$ component of Fourier space of $f(x, y)$.  
\begin{equation}
V_{m, n} = \iint ^{\frac{1}{2}} _{-\frac{1}{2}} f(x, y)\Psi _{mn}(x, y) dxdy
\end{equation}

Thus, the convolution superoperator, ${\Hat{\Hat B}}$, between ${\mathcal F}[f(\tilde{x},\tilde{y})]$ and 
${\mathcal F}[M_{mn}]$ is defined as  
\begin{eqnarray}
{\mathcal F}[f(\tilde{x},\tilde{y}) M] = {\mathcal F}[f(\tilde{x},\tilde{y})] * {\mathcal F}[M] \rightarrow \Hat{\Hat{B}} \Hat{\Hat{M}}.
\end{eqnarray}

Here, it is worth to note that 
the computational effort of the convolution in Eq.(\ref{2Dconv}) is proportional to $\mathcal{O}( m^2 \times n^2 )$, so this calculation becomes very huge. 
To avoid this difficulty, one can employ the transform method utilizing fast-Fourier Transform (FFT) in oder to diminish the computational time~\cite{FS}.
The computational effort of the transform method is proportional to $\mathcal{O}( m\log m \times n\log n )$. 
In the transform method,
one calculate the elaborated formula:
\begin{eqnarray}
{\mathcal F}[{\mathcal F}^{-1}[f_{mn}] \times {\mathcal F}^{-1}[M_{mn}]].
\end{eqnarray}
By the virtue of fast Fourier Transform(FFT) algorithm,
one can reduce the number of multiplication to calculate convolution.
Unfortunately, we could not find the superoperator for the convolution,
then, we implemented the direct calculation of convolution by using a parallel programming technique of the Message Passing Interface(MPI) standard.

\section{Results and Discussion}
\subsection{Comparison with the result of the second-order cumulant expansion}
\begin{center}
\begin{figure}[htbp]
\includegraphics[scale=0.35]{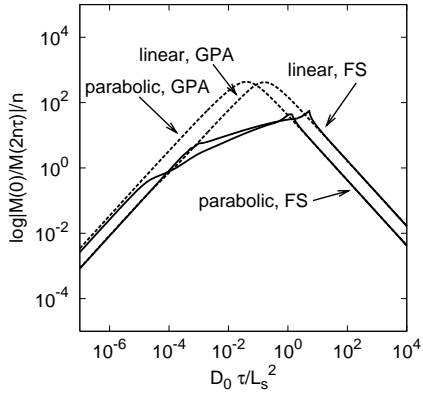}
\caption{Comparison of numerical solution of 1D Bloch-Torrey equation by 
the Fourier spectrum(FS) method, with the results of the second-order cumulant 
expansion(gaussain phase approximation;GPA).   The calculation was performed for the cases of linear($f(x)=x$)
and parabolic($f(x)=4x^2-0.5$) magnetic field grandients.
The lowest eigen modes up to 64th were taken into account.
}
\label{gpa}
\end{figure}
\end{center}

Fig.\ref{gpa} shows a comparison of the numerical solution of Bloch-Torrey equation by the Fourier spectrum method, with the results of the second-order cumulant expansion(so-called gaussian phase approximation).
The calculation was performed for the cases of a linear($f(x)=x$) and parabolic($f(x)=4x^2-0.5$) magnetic field gradients in one-dimensional. 
According to Axelrod and Sen,
the second cumulant expansion become a good approximation for both the short time and motionaly averaging regimes\cite{Sen}.
For the linear case, we obtained the exactly same behavior on the result of Fourier spectral method as the result by Axelrod and Sen~\cite{Sen}.
As for the localized regime,
the relaxation exponent indicates strong dependence on the functional shape of a magnet field gradient.

\subsection{Effect of Dimensionality of Diffusion}
\begin{center}
\begin{figure}[htbp]
\includegraphics[scale=0.35]{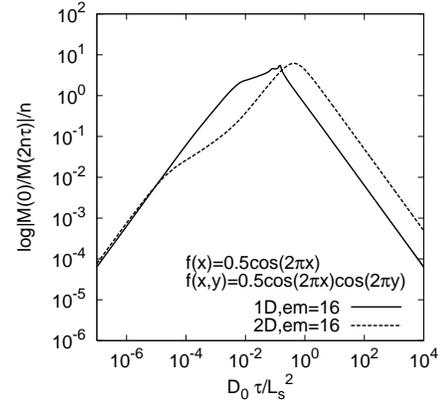}
\caption{Effect of dimensionality of diffusion on decay exponent of CPMG spin echo experiments.
The eigen modes up to 16th is taken into account and the value of $\tilde{\gamma}$ is set at 10.}
\label{1D_2D}
\end{figure}
\end{center}

A sinusoidal function may be a good candidate as a first step, for 
the expression of internal magnetic field gradient in a crystal.
For example, the Peierls potential is often approximated by the sinusoidal. 
Fig.\ref{1D_2D} shows the decay exponentials for the 8th CPMG echos 
under the condition of the sinusoidal gradients, shown in the figure, 
as a function of dimensionless diffusion coefficient, $\tilde{D_0}$
($\tilde{D_0}=\tilde{D_{x0}}$ for one-dimensional cases and 
$\tilde{D_0}=(\tilde{D_{x0}}^2 + \tilde{D_{y0}}^2)^{1/2}$ for two-dimensional cases).  
We took into account the lowest eigen modes up to 16th and 
used the $\tilde{\gamma}$ value of 10.
For the so-called localization regime, 
the lower attenuation of decay exponential was observed 
for the 2D diffusion case than that for the 1D case.
It is also found that the dimensionality of diffusion is insensitive 
both at the short time and at the motional averaging regimes.
At the motional averaging regimes,
although a slight increase of echo attenuation occurs on 2D cases, 
the decay exponent as a function of $\tilde{D_0}$ shows the same slope 
irrespective of dimensionality of diffusion.

\subsection{Effect of Anisotropic Diffusion in a 2D lattice}
\begin{center}
\begin{figure}[htbp]
\includegraphics[scale=0.35]{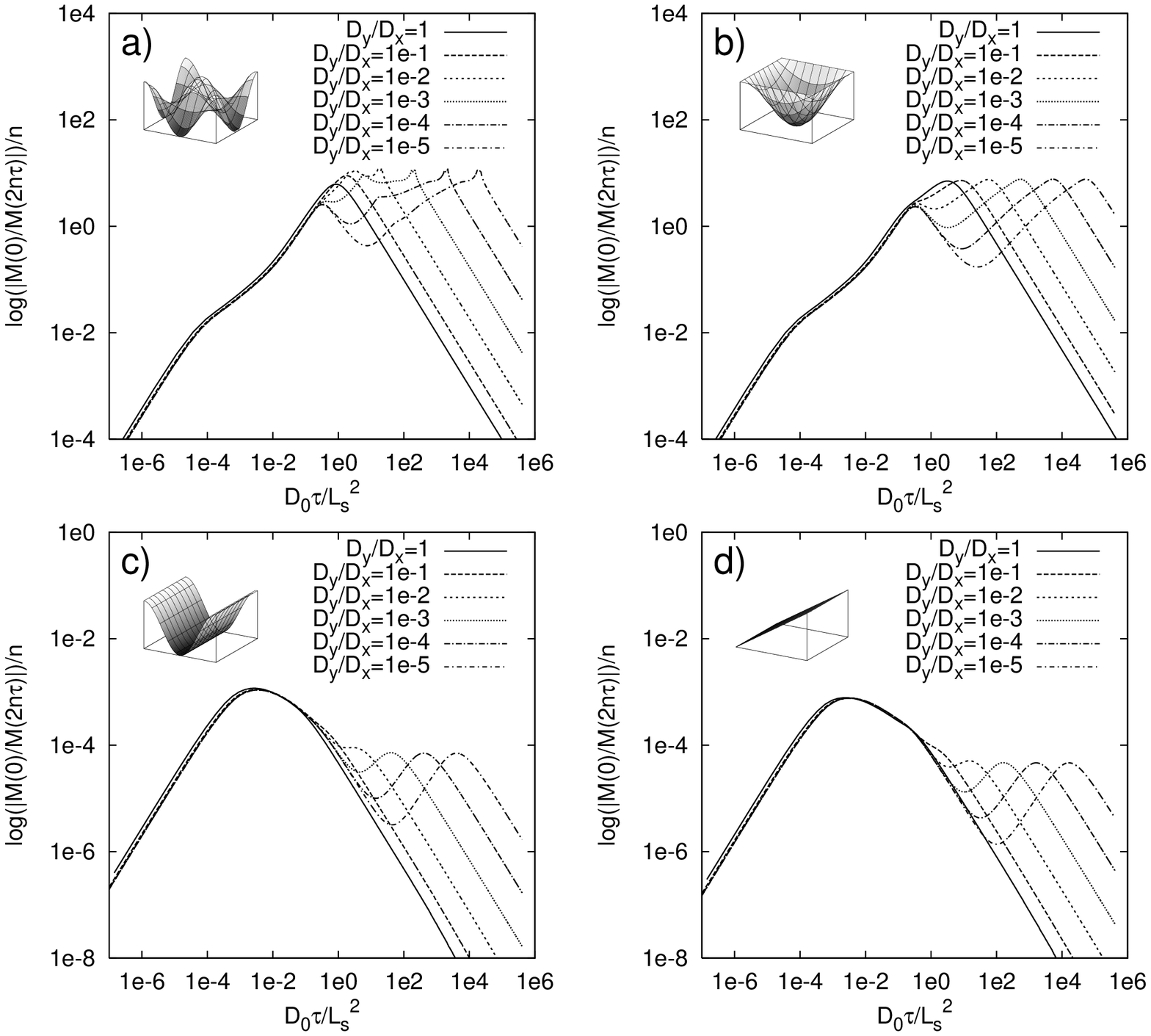}
\caption{Effect of anisotropic diffusion on decay exponent of CPMG spin echo experiments. a) $f(x,y)=0.5\cos(2\pi x)\cos(2\pi y)$, b) $f(x,y)=0.5-\cos(\pi x)\cos(\pi y)$, c) $f(x,y)=-0.5\cos(2\pi x)$, and d) $f(x,y)=x$.}
\label{aniso}
\end{figure}
\end{center}

Now let us look at a situation 
that a system of interest shows the anisotropic diffusion from internal and/or 
external reasons.
For example, 
putting the sample in the external electric field corresponds to the latter case.
Fig.\ref{aniso} shows the effect of anisotropy of diffusion on CPMG decay exponent.
One observe a double maximum profile of relaxation exponent for diffusion with an anisotropy of over 200.
Two peaks of the relaxation exponent have some interesting features.

First, 
while the main peak around the $\tilde{D_0}$ value of unity 
does not nearly move its position with respect to changes in anisotropy,
the second peak moves to positions with larger $\tilde{D_0}$ values 
proportional to the anisotropy.
Second, 
the ratio of the relaxation exponent at two maxima is 
almost constant with respect to changes in the anisotropy of diffusion, 
but the ratio and the shape of peak are strongly dependent to functional form of the field gradient.
Third,
the differences in dimensionality of diffusion show much larger effect on relaxation exponent 
than functional form of MFG.
By comparing Fig.\ref{aniso}a) and b) with c) and d), 
one can realize that 
the relaxation exponents for two-dimensional MFG are much larger than those for
one-dimensional for the case of isotropic diffusion($D_{x0}/ D_{y0} = 1$).

From the above results, 
it would be possible to determine the functional form of MFG 
by transverse relaxation time measurements at localization regime
when the anisotropy of diffusion is large.
If the MFG is originated from internal one, 
for example, crystalline periodicity, Peierls potential~\cite{PP1,PP2,PP3}, Friedel oscillation~\cite{Yu}, and so on, 
transverse relaxation time measurements will become a novel tool for
determination of the functional form of these internal properties.

\section{Conclusion}
We have explored diffusion and internal local magnetic field gradients in solids 
by nuclear magnetic resonance(NMR) transverse relaxation time measurements. 
The Fourier-spectrum method was employed in order to solve the Bloch-Torrey equation
with arbitrary magnetic field gradients in one- and two-dimensional lattice restrictions.
If there are no phase transitions or deformations of samples for
a temperature region interested, 
variable temperature CPMG measurements with a fixed echo time have 
a possibility for determination of dimensionality of diffusion processes 
found in the samples. 
The method will be applicable to several research areas such as 
transport phenomena of charge carriers in ion conducting materials,
fluid dynamics in low dimensional porus materials, and
dynamics of low dimensional elementary excitations in solids.

\section*{Acknowledgments}
This work is supported 
by Ministry of Education, Culture, Sports, Science and Technology(Japan)
through a Grant-in-Aid for science research(No.16685012).

\end{document}